\documentclass{iaus}
\usepackage{graphicx}
\title[S264-p:44 -- Sounding stellar cycles with Kepler] 
{Sounding stellar cycles with Kepler -- preliminary results from ground-based chromospheric activity measurements
\footnote{Based on observations made with the Nordic Optical Telescope, operated
on the island of La Palma jointly by Denmark, Finland, Iceland,
Norway, and Sweden, in the Spanish Observatorio del Roque de los
Muchachos of the Instituto de Astrof{\'i}sica de Canarias.}}

\author[Karoff et al.]   
{C. Karoff$^1$, T.~S.~Metcalfe$^2$, W.~J.~Chaplin$^1$, S.~Frandsen$^3$, F.~Grundahl$^3$, H.~Kjeldsen$^3$, D.~Buzasi$^4$, T.~Arentoft$^3$ \&~J.~Christensen-Dalsgaard$^3$, }

\affiliation{
$^1$School of Physics and Astronomy, University of Birmingham, Edgbaston,~Birmingham~B15~2TT, UK \\ 
$^{2}$High Altitude Observatory and Scientific Computing Division, NCAR, 
PO~Box~3000,~Boulder,~CO~80307,~USA\\
$^{3}$Department of Physics and Astronomy, Aarhus~University,~DK-8000~Aarhus~C,~Denmark\\
$^{4}$Eureka Scientific, Inc., 2452~Delmer~Street~Suite~100,~Oakland,~CA ~94602,~USA\\
email: {\tt karoff@bison.ph.bham.ac.uk}
}

\pubyear{2009}
\volume{264}  
\jname{Solar and Stellar Variability Impact on Earth and Planets}
\editors{A.~H.~Andrei, A.~Kosovichev, J-P.~Rozelot, eds.}
\begin{document}

\maketitle

\begin{abstract}
Due to its unique long-term coverage and high photometric precision, observations from the Kepler asteroseismic investigation will provide us with the possibility to sound stellar cycles in a number of solar-type stars with asteroseismology. By comparing these measurements with conventional ground-based chromospheric activity measurements we might be able to increase our understanding of the relation between the chromospheric changes and the changes in the eigenmodes.

In parallel with the Kepler observations we have therefore started a programme at the Nordic Optical Telescope to observe and monitor chromospheric activity in the stars that are most likely to be selected for observations for the whole satellite mission. The ground-based observations presented here can be used both to guide the selection of the special Kepler targets and as the first step in a monitoring programme for stellar cycles. Also, the chromospheric activity measurements obtained from the ground-based observations can be compared with stellar parameters such as ages and rotation in order to improve stellar evolution models.
\keywords{Sun: activity, Sun: oscillations, stars: activity, stars: oscillations}
\end{abstract}

\firstsection 
\section{Introduction}
During recent years there has been a growing interest in understanding stellar cycles in general and the solar cycle in particular. The reason for this is partly the indications that the solar activity cycle could cause climate change (see e.g. Svensmark \& Friis-Christensen 1997) and partly that solar cycle 24 now seems to be around two years delayed compared to theoretical predictions (Zimmerman 2009).

The ``Sounding stellar cycles with Kepler'' programme will monitor stellar cycles from the amplitude and frequency shifts of the oscillation modes observed in these stars.  By comparing these measurements with conventional ground-based chromospheric activity measurements we might be able to increase our understanding of the relation between the chromospheric changes and the changes in the eigenmodes. Asteroseismic analysis of the Kepler observations also allows us to test the hypothesis put forward by B{\"o}hm-Vitense (2007) -- that stellar cycles in active stars are generated by the gradient in the rotation rate close to the surface, whereas the stellar cycles in inactive stars are generated by the gradient in the rotation rate at the base of the convection zone. This can be done as the asteroseismic analysis of the Kepler observations will allow us to measure the depth of the convection zone and give estimates of the internal rotation profile. The details of the ``Sounding stellar cycles with Kepler'' programme are described by Karoff et al. (2009). 

The ground-based chromospheric activity measurements can be used not only to compare the changes on the stellar surfaces to those in the interior over the stellar cycles and to test stellar dynamo models, but also to calibrate rotation-activity-age relations. This can be done as asteroseismology offers a unique possibility to measure ages of solar-type field stars to a precision of 5--10\% (Christensen-Dalsgaard et al. 2007) and because asteroseismology provides the possibility to measure the inclination of the stellar rotation axis so that the spectroscopic measured $v$sin$i$ can be converted into equatorial surface rotation rates (Gizon \& Solanki 2003). Also, the first ground-based chromospheric activity measurements that we present here on the potential candidates  for observations for the whole mission can be used to rate the different candidates in order to ensure that both active and inactive stars are observed.
\begin{figure}[t]
\begin{center}
 \includegraphics[width=3.4in]{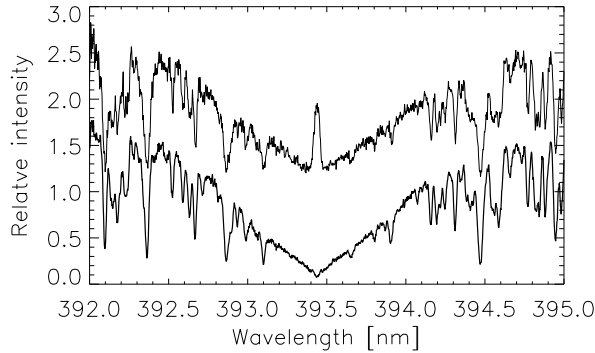} 
 \caption{Two spectra illustrating stars with and without chromospheric emission. The wavelength region covers the region around the Ca~{\sc ii} K line. The upper star (SAO 48081) clearly shows emission in the middle of the Ca~{\sc ii} K absorption line which proves that this star contains a high level of chromospheric activity. The lower star (SAO 31656) shows no emission which indicates that this star is inactive. The relative intensities have been normalized to the mean intensity in the echelle order and the spectrum for SAO 48081 has been shifted one unit upwards. No radial velocities have been calculated for the stars, but the spectra have been arbitrary aligned.}
   \label{fig1}
\end{center}
\end{figure}

\section{Observations}
During the summer of 2009 we have measured chromospheric activity for 40 of the stars that have also been observed for one month in the first part of the Kepler survey phase and are therefore candidates for being selected for observations during the whole mission. These stars were selected as described by Karoff et al. (2009) -- i.e.  as the brightest stars on the cool end of the main sequence. The observations were performed with the high-resolution FIbre-fed Echelle Spectrograph mounted on the 2.6-m Nordic Optical Telescope (Frandsen \& Lindberg 2000). One echelle spectrum with a resolution of 25000 was obtained for each star with an integration time of 420 sec. The raw spectra were reduced and wavelength calibrated using the software package $FIEStool$. Examples of reduced spectra for an active and an inactive star are shown in Fig.~1. 

In order to measure the chromospheric activity we first measure the dimensionless $S$ index as described by Hall et al. (2007) by measuring the flux in four 1-{\AA}-wide bandpasses -- two in the Ca~{\sc ii} H \& K line cores and two in the continuum away from the Ca~{\sc ii} H \& K lines. During the observations we started each night with observing the star HD~157214 which is part of the HK project at Lowell Observatory (Hall et al. 2007). It has therefore been possible to calibrate our measured $S$ indices to the $S$ indices measured at Lowell Observatory using the spectrum and $S$ index of HD~157214 measured at Lowell Observatory around the same time as our observations and kindly provided to us by J.~C.~Hall. We also plan to transform our measured $S$ indices into chromospheric emission fractions $R'_{{\rm HK}}$ and  absolute flux $\mathcal{F}_{{\rm HK}}$ (see Hall et al. 2007 for details), but this will only be done when the Kepler observations have provided us with precise asteroseismic measurements of the stellar effective temperatures.
\begin{figure}[t]
\begin{center}
 \includegraphics[width=3.4in]{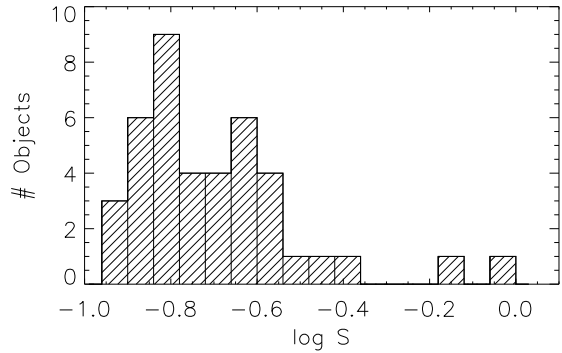} 
 \caption{Histograms of the $S$ indices for the 40 Kepler targets observed in the summer of 2009 with FIES at the NOT as part of the ``Sounding stellar cycles with Kepler'' program. The figure nicely shows the separation of the observed stars into active stars with log $S$ around $-0.6$ and inactive stars with log $S$ around $-0.8$.}
   \label{fig2}
\end{center}
\end{figure}

Figure~2 shows a histogram of the measured $S$ indices. The stars separate into a nice bifurcated distribution known as the Vaughan--Preston gap (Vaughan \& Preston 1980) and the goal for the selection of stars for observations during the whole mission is that these stars will also reflect this distribution.


\begin{thebibliography}{}
\bibitem[B{\"o}hm-Vitense(2007)]{2007ApJ...657..486B} B{\"o}hm-Vitense, E.\ 2007, ApJ, 657, 486 
\bibitem[Christensen-Dalsgaard et al.(2007)]{2007CoAst.150..350C} Christensen-Dalsgaard, J., Arentoft, T., Brown, T.~M., Gilliland, R.~L., Kjeldsen, H., Borucki, W.~J., \& Koch, D.\ 2007, Communications in Asteroseismology, 150, 350
\bibitem[Frandsen \& Lindberg(2000)]{2000mons.proc..163F} Frandsen, S., \& Lindberg, B.\ 2000, The Third MONS Workshop: Science Preparation and Target Selection, 163
\bibitem[Gizon \& Solanki(2003)]{2003ApJ...589.1009G} Gizon, L., \& Solanki, S.~K.\ 2003, ApJ, 589, 1009 
\bibitem[Hall et al.(2007)]{2007AJ....133..862H} Hall, J.~C., Lockwood, G.~W., \& Skiff, B.~A.\ 2007, AJ, 133, 862
\bibitem[Karoff et al.(2009)]{2009MNRAS.tmp.1226K} Karoff, C., Metcalfe, 
T.~S., Chaplin, W.~J., Elsworth, Y., Kjeldsen, H., Arentoft, T., \& Buzasi, D.\ 2009, MNRAS, 1226 
\bibitem[Svensmark \& Friis-Christensen(1997)]{1997JATP...59.1225S} Svensmark, H., \& Friis-Christensen, E.\ 1997, Journal of Atmospheric and Terrestrial Physics, 59, 1225
\bibitem[Vaughan \& Preston(1980)]{1980PASP...92..385V} Vaughan, A.~H., \& Preston, G.~W.\ 1980, PASP, 92, 385 
\bibitem[Zimmerman(2009)]{zimmerman} Zimmerman R.\ 2009, ScienceNOW Daily News, 8 May
\end{thebibliography}
\end{document}